\documentclass{article}

\usepackage[a4paper, total={6in, 8in}]{geometry}
\usepackage[utf8]{inputenc}
\usepackage{amsmath}
\usepackage{amssymb}
\usepackage{hyperref}
\usepackage{makecell}
\usepackage{lscape}
\usepackage{bbold}
\usepackage{graphicx}
\usepackage{changepage}
\usepackage{titlesec}
\usepackage{tabularx}
\usepackage{color, colortbl}

\usepackage{soul}
\usepackage{float}
\usepackage{authblk}
\usepackage[table]{xcolor}

\usepackage{apalike}
\makeatletter
\def\@cite#1#2{#1\if@tempswa , #2\fi}
\makeatother

\makeatother

\definecolor{Gray}{gray}{0.9}

\titleformat{\paragraph}
{\normalfont\normalsize\bfseries}{\theparagraph}{1em}{}
\titlespacing*{\paragraph}
{0pt}{1.25ex plus 1ex minus .2ex}{1.5ex plus .2ex}

\title{Comparative analysis of the yields of dicentrics and chromosomal translocations}

\author[*1]{Dorota Młynarczyk}
\author[1,2]{Pedro Puig}
\author[3]{Joan F. Barquinero}
\author[4]{Carmen Armero}
\author[5]{Virgilio Gómez-Rubio}

\affil[1]{Departament de Matemàtiques, Universitat Autònoma de Barcelona, Bellaterra (Spain)}
\affil[2]{Centre de Recerca Matemàtica, Bellaterra (Spain)}
\affil[3]{Departament de Biologia Animal, Biologia Vegetal i Ecologia, Universitat Autònoma de Barcelona,
Bellaterra (Spain)}
\affil[4]{Departament d’Estadística i Investigació Operativa, Universitat de València, València (Spain)}
\affil[5]{Department of Mathematics, School of Industrial Engineering, Universidad de Castilla-La Mancha, Albacete (Spain)}

\affil[*]{Correspondence: dorotaanna.mlynarczyk@uab.cat}
\date{}
\begin{document}
\maketitle
\section*{Abstract}
\subsection*{Purpose}
Chromosomal dicentrics and translocations are commonly employed as biomarkers to estimate radiation doses. The main goal of this article is to perform a comparative analysis of yields of both types of aberrations. The objective is to determine if there are relevant distinctions between both yields, allowing for a comprehensive assessment of their respective suitability and accuracy in the estimation of radiation doses.
\subsection*{Materials and Methods}
The analysis involved data from a partial-radiation simulation study with the calibration data obtained through two scoring methods: conventional and PAINT modified. Subsequently, a Bayesian bivariate zero-inflated Poisson model was employed to compare the posterior marginal density of the mean of dicentrics and translocations and assess the differences between them.
\subsection*{Results}
When employing the conventional method of scoring, the findings indicate that there is no notable disparity between the yield of observed translocations and dicentrics. However, when utilizing the PAINT modified method, a notable discrepancy is observed for higher doses, indicating a relevant difference in the mean number of the two types aberrations.
\subsection*{Conclusions}
The choice of scoring method significantly influences the analysis of radiation-induced aberrations, especially when distinguishing between complex and simple chromosomal formations. Further research and analysis are necessary to gain a deeper understanding of the factors and mechanisms impacting the formation of dicentrics and translocations.
\subsection*{Keywords} bivariate
zero-inflated Poisson model, FISH-painting, PAINT and conventional nomenclatures, radiation-induced chromosome aberrations
\section{Introduction}
The quantitative assessment of dicentrics and translocations, two well-studied chromosomal aberrations induced by exposure to ionizing radiation, plays a pivotal role in radiation biology. Sometimes it is assumed that yields for these two types of aberrations are very similar, but concerns may arise regarding their potential inequality, highlighting the complex mechanisms that influence these chromosomal changes in response to varying radiation doses. The current radiation biodosimetry manual (\cite{IAEA2011}) recommends that each laboratory should develop its own fitting dose-response curves for translocations, but, in practice, in situations where a laboratory lacks a specific translocations curve, it is occasionally acceptable to use the calibration curve for dicentrics (\cite{barquinero2017}). Furthermore, the analysis of aberrations may also be influenced by the detection technique employed and the terminology used to characterize these alterations (\cite{barquinero1999}). Thus, the objective of this article is to compare yields for dicentrics and translocations obtained using different scoring systems, with the aim of evaluating the technique's influence on the disparities between these yields. While some authors have previously tackled this issue (\cite{barquinero1999}; \cite{lindholm1998}), our current article emphasizes a new mathematical approach.

Radiation-induced chromosome exchange aberrations produced in lymphocytes in G0 (quiescent) stage have been classified as symmetrical or asymmetrical (i.e reciprocal translocations or dicentrics) (\cite{savage1982}). For biological dosimetry purposes and in solid stained metaphases, symmetrical exchanges are difficult to detect, unless the resulting chromosomes are markedly different from the normal karyotype. For this reason, initially biological dosimetry was based on the detection of dicentrics or dicentrics plus rings (\cite{IAEA2001}).  From the visualization of a solid stained dicentric plus its corresponding acentric fragment it was logical to infer that this exchange resulted to the interaction of two broken chromosomes, and by association the same for symmetric exchanges like translocations. The introduction of fluorescence in situ hybridization techniques (FISH) to detect whole chromosomes (chromosome-painting) allowed an easy detection of translocations (\cite{lucas1992}). However, with the same technique, it was evident that radiation-induced chromosome exchanges could be formed by the interaction of more than two breaks (\cite{cremer1990}; \cite{schmid1992}) and exchanges were classified between simple and complex. The former involve two breaks and the latter involving three or more breaks in two or more chromosomes (\cite{savage1994}).

The first calibration curves produced by chromosome-painting used the terms translocation and dicentric similarly to that was done when analyzing solid stained metaphases (\cite{lucas1992,bauchinger1993,fernandez1995}), i.e. distinguishing aberrations: translocation(\textit{t}), dicentric(\textit{dic}), ring(\textit{r}), insertion(\textit{ins}), and additional acentric(\textit{ace}) fragment (\cite{iscn1985}). This system, hereafter referred to as the conventional system, may also identify complete or incomplete dicentrics and translocations, along with centric or acentric rings. However, the presence of complex exchanges mainly at higher doses led the development of specific nomenclatures to describe the radiation-induced exchanges. Initially, two highly different nomenclatures were proposed: the nomenclature with the acronym PAINT (from Protocol for Aberration Identification and Nomenclature Terminology; \cite{tucker1995}) was purely descriptive of each abnormal piece present in the metaphase, without cross-reference to other aberrant pieces in the same metaphase; and the CAB terminology (from Chromosomes-Arms-Breaks; \cite{cornforth2001}) or  the so-called S\&S system (\cite{savage1994}) that proposed a code with numerals and letters that refer to the number of abnormal pieces observed and how common the exchange was expected to be. Understanding complex aberrations is greatly aided by the CAB terminology, which focused on the mechanical aspects of exchange formation. However, in daily practice the use of the CAB vocabulary was not easy to handle, and currently the most widely used method to score the chromosome exchanges using FISH is to describe each abnormal metaphase as a unit using the PAINT nomenclature in a slightly modified way that allows to infer the mechanistic aspects of exchange formation and allows classifying between simple and complex aberrations (\cite{knehr1998}). Specifically, within the PAINT modified system, terms like 'apparently simple dicentric/translocation' (ASD/AST) are employed to describe aberrations involving bicolored chromosomes with single-color junctions each (\cite{tucker1995}), which may be the result of complex, undetected aberrations.

\begin{figure}[ht]\centering
\includegraphics[scale=0.4]{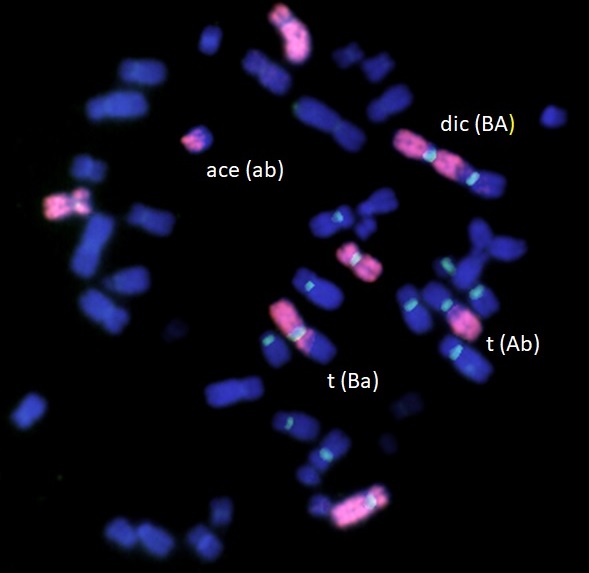}
\caption{\label{fig:FISH} 
An example of a metaphase illustrating FISH-based chromosome ‘painting’. In red painted chromosomes \#1, \#4, and \#11, in blue counterstained chromosomes.}
\end{figure}

Despite the existence of these nomenclatures, some biological dosimetry laboratories still rely on conventional terminology to describe chromosomal alterations. Therefore, the question is how different scoring systems (conventional vs. PAINT modified) influence the analysis of yields of aberrations, and whether these can lead to misleading interpretations of the results. To illustrate the problem, let's consider an example from Figure \ref{fig:FISH}. The metaphase contains four aberrant pieces with painted material involved. The metaphase would be scored as: \textit{dic}(BA), \textit{t}(Ba), \textit{ace}(ab), \textit{t}(Ab). These aberrations could be scored by combining aberrations that imply the minimum number of breaks, i.e., \textit{dic}(BA) with \textit{ace}(ab), which is an ASD or a dicentric, and \textit{t}(Ba) with \textit{t}(Ab), which is an AST or a translocation. However, one could also combine \textit{dic}(BA) with \textit{t}(Ab), resulting in a complex aberration (neither ASD nor AST), and it will be counted as one dicentric plus one translocation in the conventional nomenclature. Similarly, combining \textit{t}(Ba) with \textit{ace}(ab) results in a complex aberration, which will be counted as a translocation.

For the purpose of this research, we decided to utilize the data obtained from the study conducted by \cite{duran2002} (2002). The study involved a blood sample taken from a healthy 32-year-old male that were exposed to X-rays at radiation doses of 2, 3, 4, and 5 Gy within a laboratory environment. Subsequent to the exposure, the samples underwent the FISH protocol with chromosomes 1, 4, and 11 being painted. The scoring of chromosomal aberrations was carried out using both the conventional and PAINT modified methods. This implies that we have counts of dicentrics and translocations via the conventional scoring approach, as well as counts of ASD and AST using the PAINT modified approach for each cell, what enables a comparison between the results obtained from both methods.

\cite{duran2002} (2002) studied specifically partial body radiation exposures. This type of exposure occurs when only a specific region or part of the body is subjected to radiation, such as during targeted radiation therapy or accidental exposure to a localized radiation source. To replicate this situation experimentally, various proportions of irradiated and non-irradiated blood were mixed together: 0.875, 0.75, 0.5, 0.25, and 0.125. Furthermore, the analysis also considered the impact of total irradiation by including samples with a proportion of 1 (without the addition of non-irradiated blood to the samples). Subsequently, these mixed samples were carefully scored using both considered terminologies.

\begin{table}[H]
\centering
\resizebox{\textwidth}{!}{%
\begin{tabular}{ll|cccccc|cccccc}
  \hline
&dose (Gy) & 2 & 2 & 2 & 2 & 2 & 2 & 3 & 3 & 3 & 3 & 3 & 3 \\ 
&proportion of irradiated cells $p$ & 1 & 0.875 & 0.75 & 0.50 & 0.25 & 0.125 & 1 & 0.875 & 0.75 & 0.50 & 0.25 & 0.125  \\ 

& number of cells & 525 & 974 & 1551 & 1322 & 1516 & 1322 & 394 & 509 & 824 & 1009 & 1070 & 1096  \\ 
\hline 
conventional & mean dicentrics & 0.12 & 0.07 & 0.06 & 0.04 & 0.02 & 0.01 & 0.21 & 0.22 & 0.10 & 0.08 & 0.03 & 0.02  \\ 
& mean translocations & 0.11 & 0.08 & 0.06 & 0.05 & 0.02 & 0.01 & 0.27 & 0.21 & 0.10 & 0.10 & 0.03 & 0.04 \\ 
  \hline
PAINT mod. & mean ASD & 0.10 & 0.07 & 0.06 & 0.04 & 0.02 & 0.01 & 0.16 & 0.21 & 0.10 & 0.07 & 0.03 & 0.02  \\ 
& mean AST & 0.08 & 0.07 & 0.05 & 0.04 & 0.02 & 0.01 & 0.20 & 0.17 & 0.08 & 0.07 & 0.02 & 0.03 \\ 

\hline \hline
& dose (Gy) & 4 & 4 & 4 & 4 & 4 & 4 & 5 & 5 & 5 & 5 & 5 & 5 \\ 
& proportion of irradiated cells $p$ & 1 & 0.875 & 0.75 & 0.50 & 0.25 & 0.125 & 1 & 0.875 & 0.75 & 0.50 & 0.25 & 0.125 \\ 
& number of cells & 250 & 463 & 504 & 775 & 1035 & 1040 & 133 & 374 & 553 & 523 & 890 & 1260 \\ 

\hline
conventional & mean dicentrics & 0.48 & 0.23 & 0.16 & 0.10 & 0.04 & 0.02 & 0.47 & 0.35 & 0.20 & 0.11 & 0.06 & 0.02 \\
& mean translocations & 0.50 & 0.20 & 0.19 & 0.12 & 0.04 & 0.02 & 0.59 & 0.30 & 0.18 & 0.13 & 0.06 & 0.02 \\
\hline
PAINT mod. & mean ASD &0.38 & 0.20 & 0.14 & 0.08 & 0.04 & 0.01 & 0.33 & 0.27 & 0.18 & 0.08 & 0.05 & 0.02 \\ 
& mean AST & 0.33 & 0.16 & 0.14 & 0.08 & 0.03 & 0.02 & 0.42 & 0.20 & 0.13 & 0.09 & 0.04 & 0.01 \\ 
\hline
\end{tabular}}
\caption{Mean number of aberrations for each dose and dilution $p$, utilizing both the conventional and PAINT mod. scoring systems.}
\label{tab:dataset}
\end{table}

Table \ref{tab:dataset} displays the mean number of aberrations for each dose and dilution $p$ (proportion of irradiated cells), utilizing both the conventional and PAINT mod. nomenclature. In total, 20,332 cells were scored, with the lowest count of 3,733 at a dose of 5Gy and of 1,302 for proportion $p=1$. This table highlights relevant differences between the two scoring systems under consideration as evident in the variations in the mean numbers of dicentrics/ASD and translocations/AST. The raw data can be found in the supplementary material of this article.

\section{Materials and methods} \label{sec:stat}
From a statistical standpoint, the bivariate Poisson model seems to be a suitable choice when working with both dicentrics and translocations, as it allows for joint analysis and inference about the underlying processes. This model assumes that both count variables are generated by two different Poisson processes, with a specific correlation term between them. Moreover, the bivariate Poisson model can be extended to cover scenarios with partial body exposures. The experimental design, which combines irradiated and non-irradiated samples, results in a significant number of zero counts in the observed data, which can be effectively handled by a bivariate zero-inflated Poisson model. While bivariate Poisson models have been widely employed for analyzing sports and healthcare data (\cite{karlis2003}), their application in the field of biodosimetry has been largely unexplored. 
Furthermore, the bivariate approach utilized in this study for analyzing dicentrics and translocations can also be applied to analyze dicentrics and rings or dicentrics and foci. However, it would be more difficult in the case of foci due to their significant dependence on the amount of time that has passed following irradiation (\cite{mlynarczyk2022}). 

To begin the probabilistic formulation of the sampling model, let's clarify that we are defining a single model for both scoring techniques. Therefore, in this section, we will simply refer to aberrations as dicentrics and translocations, without mentioning ASD or AST.
Assume that $N$ cells were examined. We denote the number of dicentrics $X_j$ and of translocations $Y_j$ found in cell $j$, $j=1, \ldots, N$. Then, it is assumed that $X_j$ and $Y_j$ jointly follow a conditional zero-inflated bivariate Poisson distribution, with probability function

\begin{equation}
  f_{ZIBP}(x_j,y_j\mid \omega, \lambda_{1j}, \lambda_{2j}, \lambda_{3j}) =
    \begin{cases}
      \omega f_{BP}(x_j,y_j \mid \lambda_{1j}, \lambda_{2j}, \lambda_{3j}) & (x_j,y_j) \neq (0,0)\\
      (1-\omega)+\omega f_{BP}(x_j,y_j \mid \lambda_{1j}, \lambda_{2j}, \lambda_{3j}) & (x_j,y_j)= (0,0)\\
    \end{cases}       
\end{equation}

where $(1-\omega)$ is the proportion of structural zeros (proportion of not-irradiated cells), and $f_{BP}(x_j,y_j \mid \lambda_{1j}, \lambda_{2j}, \lambda_{3j})$ is the conditional probability function of the bivariate Poisson distribution given by

\begin{equation*}
    f_{BP}(x_j,y_j \mid \lambda_{1j}, \lambda_{2j}, \lambda_{3j})=\exp(-(\lambda_{1j}+ \lambda_{2j}+ \lambda_{3j})) \frac{\lambda_{1j}^{x_j}}{x_j!} \frac{\lambda_{2j}^{y_j}}{y_j!} \sum_{k=0}^{\min(x_j,y_j)} \binom{x_j}{k} \binom{y_j}{k} k! (\frac{\lambda_{3j}}{\lambda_{1j} \lambda_{2j}})^k.
\end{equation*}

\noindent
The effect on the irradiated cells is determined by the parameters $\lambda_{1j}, \lambda_{2j}$, and $\lambda_{3j}$ in our model. In more detail, $\lambda_{1j} + \lambda_{3j}$ represents the marginal mean of the number of dicentrics, $\lambda_{2j} + \lambda_{3j}$ the marginal mean of translocations, and $\lambda_{3j}$ the covariance term between the number of dicentrics and translocations (all them for the irradiated cells). These parameters, all them greater than zero, can be modelled as functions of the radiation dose using a regression approach. The mean number of aberrations is frequently estimated in biodosimetric research using a linear or quadratic function of dose; we chose to concentrate on the quadratic model because the linear model is also included in it. We specify them as,

\begin{equation}
\begin{aligned}
\text{dicentrics} \\
\text{translocations} \\
\text{covariance} \\
\end{aligned}
\quad
\begin{aligned}[c]
    \lambda_{1j} &= \beta_{11}+ \beta_{12} \cdot d_j + \beta_{13} \cdot d^2_j, \\ 
    \lambda_{2j} &= \beta_{21}+ \beta_{22} \cdot d_j + \beta_{23} \cdot d^2_j , \\
    \lambda_{3j} &= \beta_{31},
\end{aligned}
\label{eq:def_models}
\end{equation}

\noindent
where $d_j$ denotes the radiation dose received by $j$-th cell, and $\beta_{\bullet\bullet}$ denotes the corresponding regression coefficients.  The expected number of dicentrics, including irradiated and not-irradiated cells, is $\mu_{1j}=E(X_j \mid \beta_{11}, \beta_{12}, \beta_{13} )=\omega(\lambda_{1j}+\lambda_{3j})$. Similarly for translocations $\mu_{2j}=E(Y_j \mid \beta_{21}, \beta_{22}, \beta_{23} )=\omega(\lambda_{2j}+\lambda_{3j})$.

It has been studied that some cells exposed to radiation fail to survive until the metaphase stage of the cell cycle, when dicentrics and translocations are analyzed (\cite{hall2006}). As a result, the actual proportion of irradiated cells present in the sample at the time of the analysis is lower than the initially assumed proportion. This discrepancy between both proportions poses a challenge in accurately assessing the radiation-induced chromosomal aberrations. Therefore, it is essential to account for this limitation, so researchers employ a statistical correction to estimate the true proportion of irradiated cells (\cite{pujol2016}). The survival rate, the proportion of the irradiated cells which survive until the metaphase, is described as a decreasing exponential function of the dose, 

\begin{equation*}
s(d)=\text{exp}(-\gamma d),
\label{eq:survival_index}
\end{equation*}

where $\gamma$ is the so-called survival index, $\gamma \in [0,1]$. Therefore the actual proportion $\omega$ of scored cells irradiated at dose $d$ is given by (see supplementary material for details) 

\begin{equation}
    \omega(d,p)= \frac{1}{1+\text{exp}(\gamma d)(1-p)/p},
\label{eq:omega}
\end{equation}

where $p$ is the initial proportion of the irradiated cells. Note that in the given experiment $d$ and $p$ are known, but the survival index $\gamma$ needs to be estimated. 

Let denote by $\mathbf{x}=(x_1, \ldots, x_N)$ the counts of dicentrics in cells $1,\ldots,N$, and respectively the counts of translocations by $\mathbf{y}=(y_1,\ldots, y_N)$. The vector $\boldsymbol{d}=(d_{1}, \ldots, d_{N})$ stands for the radiation doses to which each cell was exposed and $\boldsymbol{p}=(p_{1}, \ldots, p_{N})$ is the vector of initial proportion of irradiated cells in the sample from which the observation comes. Thus the likelihood of the model is given by

\begin{align*}
    L(\mathbf{x}, \mathbf{y}, \boldsymbol{d}, \boldsymbol{p} \mid \boldsymbol{\beta}, \gamma )&= \prod_{j=1}^N \Biggr( 1_{(0,0)}(x_j,y_j) \left( 1-\frac{1}{1+\text{exp}(\gamma d_j)(1-p_j)/p_j} \right) \\ &+\frac{1}{1+\text{exp}(\gamma d_j)(1-p_j)/p_j} \ \text{exp}(-(\lambda_{1j}+ \lambda_{2j}+ \lambda_{3j})) \frac{\lambda_{1j}^{x_j}}{x_j!} \frac{\lambda_{2j}^{y_j}}{y_j!} \\ & \times \sum_{k=0}^{\min(x_j,y_j)} \binom{x_j}{k} \binom{y_j}{k} k! (\frac{\lambda_{3j}}{\lambda_{1j} \lambda_{2j}})^k \Biggr),
\end{align*}

\noindent
where $\lambda_{ij}$ are defined in (\ref{eq:def_models}), and $1_{(0,0)}(x_j,y_j)$ is an indicator function that is 1 when $x_j=y_j=0$ and 0 otherwise.  Within the Bayesian framework, the main interest is in computing the posterior distribution $\pi(\boldsymbol{\beta}, \gamma \mid \mathbf{x}, \mathbf{y}, \boldsymbol{d}, \boldsymbol{p})$ of the parameters of the model, i.e. the regression coefficients $\boldsymbol{\beta}=(\beta_{11},\ldots ,\beta_{31})$ and the survival index $\gamma$, through the Bayes´ theorem 

\begin{equation*}
\pi(\boldsymbol{\beta}, \gamma \mid \mathbf{x}, \mathbf{y}, \boldsymbol{d}, \boldsymbol{p}) \propto L(\mathbf{x}, \mathbf{y}, \boldsymbol{d}, \boldsymbol{p} \mid \boldsymbol{\beta}, \gamma) \pi( \boldsymbol{\beta},\gamma),
\end{equation*}
where $\pi( \boldsymbol{\beta},\gamma)$ is the prior distribution for $( \boldsymbol{\beta},\gamma)$. We assume prior independence between $\boldsymbol{\beta}$ and $\gamma$, so $\pi( \boldsymbol{\beta},\gamma)=\pi( \boldsymbol{\beta})\pi(\gamma)$. Given the fact that $\lambda_{1j}$ and $\lambda_{2j}$ should be non-negative, the prior distribution $\pi(\boldsymbol{\beta})$ was chosen to be a non-informative uniform distribution on the interval $(0,10)$. The prior distribution for $\gamma$ was chosen to be a uniform distribution on the interval $[0,1]$ because, according to the definition, the survival index belongs to the interval $[0,1]$. Approximated samples from the posterior distribution $\pi(\boldsymbol{\beta}, \gamma \mid \mathbf{x}, \mathbf{y}, \boldsymbol{d}, \boldsymbol{p})$ were generated using Markov chain Monte Carlo methods. The JAGS software (\cite{JAGS}), in particular, was utilized with specific configurations, including the use of 2 chains, conducting 25,000 simulations, burn-in of 1000 iterations, and thinning every 25 iterations 
(the program itself is available in the supplementary material).

Note that $\lambda_{1j}$ and $\lambda_{2j}$ depend on the dose received by the cell $j$ and are determined by equations (\ref{eq:def_models}). By fixing the dose and utilizing samples from the posterior distribution, it becomes straightforward to derive the posterior distribution for the expected number of aberrations. From this point onward, we will focus only  on the irradiated cells (with the actual proportion $\omega(d, p)=1$), since we regard this as the most interesting case. For a fixed dose $d$, we can omit the subindex $j$ in equations (\ref{eq:def_models}), and denote the mean number of dicentrics of the irradiated cells for dose $d$ as $\mu_{1d}$ and of translocations as $\mu_{2d}$.
Our main goal is to discuss whether these mean values differ from each other. To do so, we can consider the posterior distribution of the difference between means $\pi(\mu_{1d}-\mu_{2d} \mid \mathbf{x}, \mathbf{y}, \boldsymbol{d}, \boldsymbol{p})$. It can be easily approximated through simulation, enabling the calculation of the posterior mean of the difference, credible intervals, and conducting graphical analysis by plotting posterior densities for each dose. We will report Highest Density Intervals (HDI), which are credible intervals designed to have a higher probability density for all values inside the interval compared to those outside (\cite{kruschke2014}).

We can additionally compare the models by assessing the probability that the mean number of dicentrics of the irradiated cells, $\mu_{1d}$, is lower than the mean number of translocations, $\mu_{2d}$, separately calculated for each dose $d$. Let's define this as hypothesis $H_1: \mu_{1d} - \mu_{2d} < 0$. Conversely, the alternative scenario will be denoted as $H_2$, with $H_2: \mu_{1d} - \mu_{2d} \geq 0$. Evaluating the ratio between the posterior probability of $H_1$ and that of $H_2$, i.e.

\begin{equation}
\frac{\pi(\mu_{1d}-\mu_{2d} <0  \mid \mathbf{x}, \mathbf{y}, \boldsymbol{d}, \boldsymbol{p})}{\pi(\mu_{1d}-\mu_{2d} \geq 0 \mid \mathbf{x}, \mathbf{y}, \boldsymbol{d}, \boldsymbol{p})}
\label{eq:Bayes_factor}
\end{equation}
may assist us in determining which situation is more likely. Given our prior assumption that both hypotheses are equally probable a priori, i.e., $P(H_1)=P(H_2)=0.5$, this resulting odds can be interpreted as the Bayes factor (\cite{kass1995}) (details can be found in supplementary material).
According to the Jeffreys scale (\cite{jeffreys1961}), a Bayes factor close to 1 suggests only 'bare mention' evidence of a difference between $\mu_{1d}$ and $\mu_{2d}$. A number greater than 3 but less than 10 is considered 'substantial' evidence. Results between 10 and 30 indicate 'strong' evidence, and between 30 and 100 are categorized as 'very strong.' A Bayes factor exceeding 100 is considered 'decisive'.

\section{Results}\label{sec:results}
\begin{figure}[ht]\centering
\includegraphics[width=\textwidth]{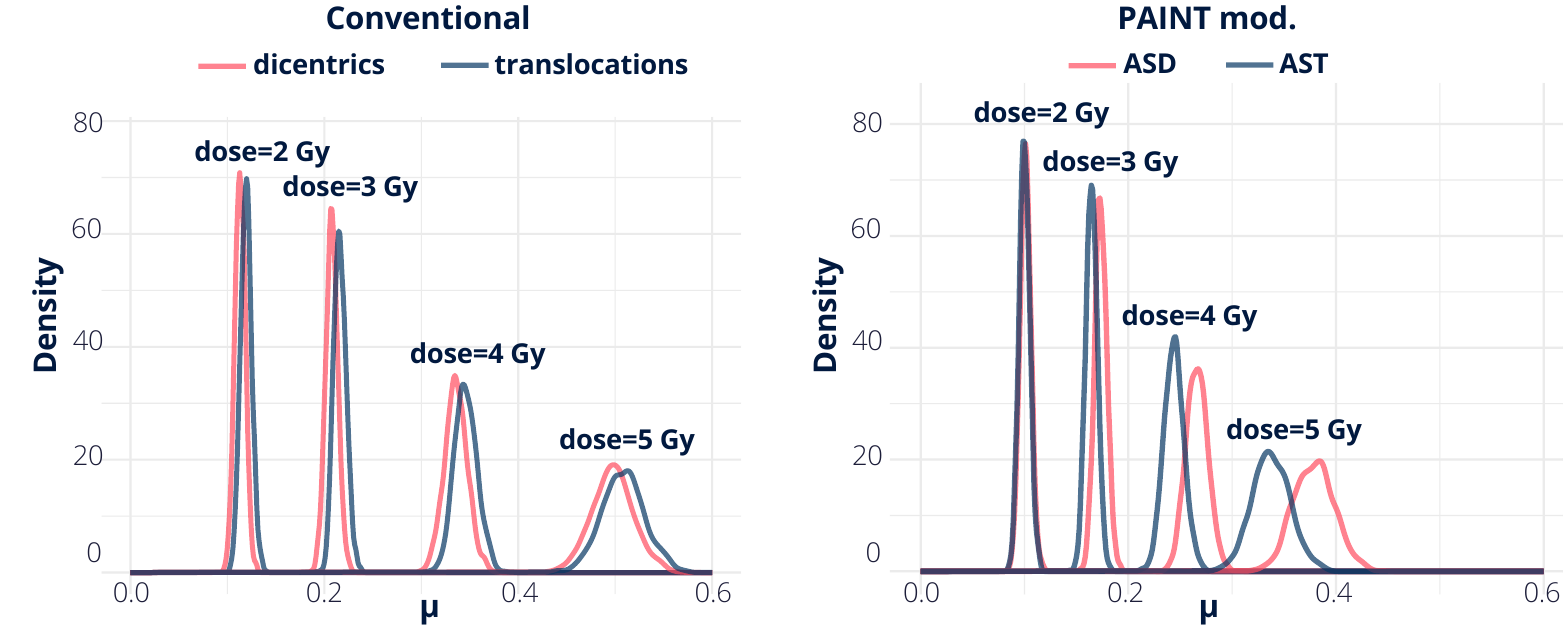}
\caption{\label{fig:lambdas_posterior} Posterior densities of the mean number of dicentrics, $\pi(\mu_{1d} \mid \mathbf{x}, \mathbf{y}, \boldsymbol{d}, \boldsymbol{p})$, and of the mean number of translocations, $\pi(\mu_{2d} \mid \mathbf{x}, \mathbf{y}, \boldsymbol{d}, \boldsymbol{p})$, given by the model for two scoring techniques.}
\end{figure}

\begin{figure}[ht]\centering
\includegraphics[width=\textwidth]{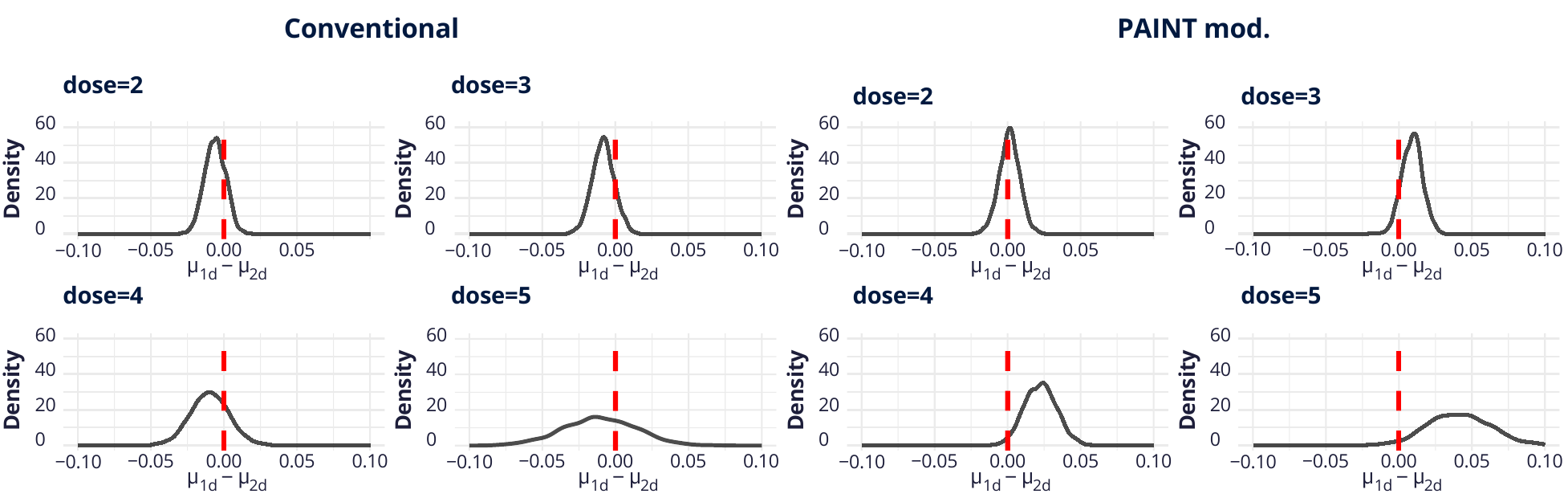}
\caption{\label{fig:diff_posterior} Posterior distribution of the difference between means of dicentrics and translocations $\pi(\mu_{1d}-\mu_{2d} \mid \mathbf{x}, \mathbf{y}, \boldsymbol{d}, \boldsymbol{p})$ given by the model for two scoring techniques. The red line is a reference line when $\mu_{1d}-\mu_{2d}=0$.}
\end{figure}

Our objective is to examine the feasibility of employing the same calibration curve for both dicentrics and translocations, which entails comparing the parameters $\mu_{1d}$ and $\mu_{2d}$ defined in Section \ref{sec:stat}. Figure \ref{fig:lambdas_posterior} depicts the posterior distribution of the means of the number of dicentrics $\mu_{1d}$ and of translocations $\mu_{1d}$ for conventional scoring method and for ASD and AST for PAINT modified technique, across different doses. Notably, this information reveals that the posterior mean values of dicentrics and translocations are relatively similar when using conventional nomenclature. The density of dicentrics of the irradiated cells is skewed to the left for each dose compared to translocations. Conversely, in the case of the PAINT modified scoring, the posterior distribution for AST is shifted towards the left. This difference becomes more pronounced as the dose increases, indicating a relevant difference between them, what can also be noticed looking at the Figure \ref{fig:diff_posterior}. This figure represents the posterior density of the difference between estimated means of dicentrics and translocations (or ASD and AST), $\pi(\mu_{1d}-\mu_{2d} \mid \mathbf{x}, \mathbf{y}, \boldsymbol{d}, \boldsymbol{p})$ for the irradiated cells at each dose and for both scoring techniques. Posterior mean and $95\%$ HDI intervals of the differences for each case can be found in the Table \ref{tab:diff_meanhdi}. All of these findings indicate that there are some distinctions in the means, especially in the case of a dose of 4Gy for the PAINT mod. method. Nevertheless, it's important to mention that more noticeable differences can be detected at higher doses.

\begin{table}[H]
\centering
\resizebox{\textwidth}{!}{%
\begin{tabular}{l|cccc|cccc}
\hline
 & \multicolumn{4}{c|}{conventional}& \multicolumn{4}{c}{PAINT mod.}  \\
 \hline
dose & 2 & 3 &4&5& 2 & 3 &4&5 \\
  \hline
mean &-0.006 & -0.008 & -0.01 & -0.011& 0.001 & 0.009 & 0.022 & 0.041  \\ 
  HDI & 
  [-0.019,0.008] & [-0.022,0.008] & [-0.035,0.017] & [-0.06,0.039] &[-0.013,0.015] & [-0.004,0.023] & [0.001,0.044] & [-0.001,0.085]   \\ 
   \hline
\end{tabular}}
\caption{Posterior mean of the difference between means of dicentrics and translocations and $95\%$ HDI for each dose and scoring technique.}
\label{tab:diff_meanhdi}
\end{table}

The results presented in Table \ref{tab:Bayes_factor} showcase the posterior odds, defined previously (\ref{eq:Bayes_factor}), as the ratio of probabilities between instances where $\mu_{1d}$ is lower than $\mu_{2d}$ and vice versa. However, for a more robust interpretation of the results, for the PAINT mod. method, the Bayes factor was computed as the reciprocal of this fraction. According to the interpretation provided by (\cite{jeffreys1961}), the results offer substantial evidence supporting the hypothesis that $\mu_{1d}$ is lower than $\mu_{2d}$ for the conventional technique, a condition generally observed, as depicted in Figure \ref{fig:lambdas_posterior}. For the PAINT technique, when the mean of ASD exceeds the mean of AST, strong evidence is found indicating that $\mu_{1d}$ is greater than or equal to $\mu_{2d}$ for doses of 4Gy and 5Gy.

\begin{table}[ht]
\centering
\begin{tabular}{rrrrr}
  \hline
dose (Gy)  & 2 & 3 & 4 &5\\ 
  \hline
conventional & 3.940 & 6.369 & 3.446 & 1.941 \\ 
PAINT mod. & 1.301 & 8.340 & 46.729 & 33.333 \\ 
   \hline
\end{tabular}
\caption{Results of Bayes factor as defined by ratio (\ref{eq:Bayes_factor}) for the conventional method of scoring and computed as the reciprocal of the fraction for the PAINT mod. method. }
\label{tab:Bayes_factor}
\end{table}

\begin{figure}[!ht]\centering
\includegraphics[width=0.7\textwidth]{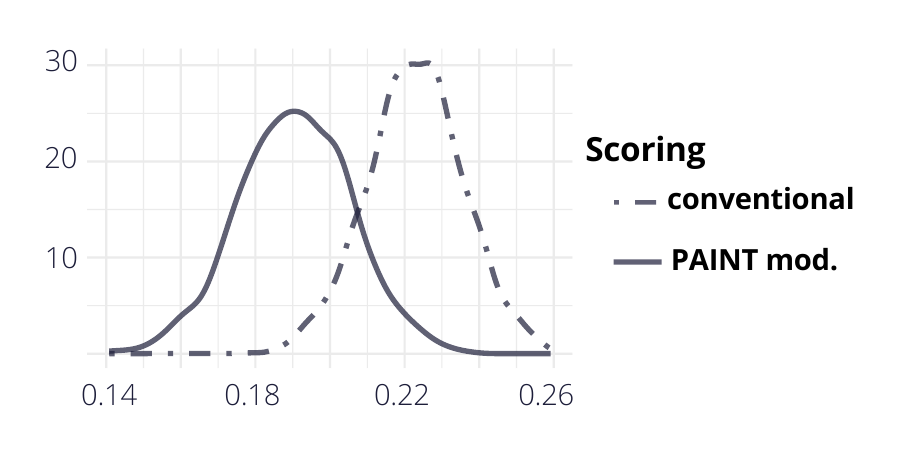}
\caption{\label{fig:posterior_gammas} Posterior density of the survival index, $\gamma$, for the two methods of scoring.}
\end{figure}

The current information about the survival index $\gamma$ of the irradiated cells is represented by the posterior distribution in Figure \ref{fig:posterior_gammas}. 
It is important to note that this survival rate, as defined in Section \ref{sec:stat}, is dose-dependent. Higher doses of radiation generally correspond to lower survival proportions, while lower doses tend to result in higher survival proportions. Additionally, the actual proportion of irradiated cells is also influenced by the initial proportion of irradiated cells as can be seen by (\ref{eq:omega}). Table \ref{tab:actual_prop} illustrates the relationship between the initial proportion of cells \textit{p}, the received dose of radiation \textit{d}, and the resulting actual proportion of surviving cells \textit{$\omega$}. These values are determined based on the mean value of the posterior density of the survival index $\gamma$. The estimated mean value of $\gamma$ for the PAINT mod. method was 0.19, while for the conventional method it was 0.22. These results indicate a significant decrease in the proportion of irradiated cells that survived until metaphase. For example, when using the conventional method with a dose of 5Gy and an initial proportion of 0.5, the estimated actual proportion of irradiated cells in the analyzed sample is approximately half of the initial proportion.

\begin{table}[H]
\centering
\begin{tabular}{|l|c|cccccc|}
\hline
& & \multicolumn{6}{c|}{Initial proportion of irradiated cells $p$} \\
Scoring method & dose $d$ (Gy) & 1 & 0.875 & 0.75 & 0.5 & 0.25 & 0.125 \\ 
  \hline
  conventional & 2 & 1 & 0.818 & 0.658 & 0.39 & 0.176 & 0.084 \\ 
 & 3 & 1 & 0.782 & 0.606 & 0.339 & 0.146 & 0.068 \\ 
 & 4 & 1 & 0.742 & 0.552 & 0.291 & 0.120 & 0.055 \\ 
 & 5 & 1 & 0.697 & 0.496 & 0.247 & 0.099 & 0.045 \\ 
   \hline
PAINT modified & 2 & 1 & 0.827 & 0.672 & 0.406 & 0.185 & 0.089 \\ 
 & 3 & 1 & 0.798 & 0.629 & 0.361 & 0.158 & 0.075 \\ 
 & 4 & 1 & 0.766 & 0.583 & 0.318 & 0.135 & 0.062 \\
& 5 & 1 & 0.730 & 0.536 & 0.278 & 0.114 & 0.052 \\
\hline
\end{tabular}
\caption{Actual proportion of irradiated cells $\omega$ estimated by the posterior mean of the survival $\gamma$ given the dose of radiation $d$ (in Gy) and the initial proportion $p$.}
\label{tab:actual_prop}
\end{table}

\section{Discussion}\label{sec:disscusion} 
As expected, the results of this study suggest that the choice of the scoring method significantly influences the analysis of the radiation-induced aberrations. However, the differences become less evident when the focus is on yields of aberrations using the same scoring technique. When using the conventional scoring method, the results suggest that there is a very small difference between the mean numbers of translocations and dicentrics observed. However, with the PAINT mod. method, a more pronounced variation in the results was found, but it's important to emphasize that our focus was solely on apparently simple aberrations. We believe that these results are primarily attributable to the fact that the conventional method does not differentiate between complex and simple chromosomal formations, whereas the PAINT mod. method does.

The prevailing understanding in past studies suggests that dicentrics and translocations have equal probability of appearing after chromosomal breaks induced by ionizing radiation (\cite{lucas1996}). However, it is important to note that there is no unanimous consensus of an equal formation ratio (1:1) and this theoretical assumption is occasionally questioned (\cite{barquinero1998}). Based on the findings from the PAINT mod. method, our results indicate that, when complex aberrations are excluded, there is a noteworthy difference in the mean number of dicentrics and translocations, particularly at higher radiation doses. This result suggests that the 1:1 ratio may not be valid; if this is the case, the results may imply that translocations would occur more frequently than dicentrics in complex aberrations.

However, another explanation of the results for PAINT mod. method, when the yield of AST is shifted left comparing to ASD, may be the limit of detection of exchanged chromosome fragments. It has been described that using FISH techniques the minimum detectable size is approximately 11 and 14 Mb for painted and unpainted material, respectively (\cite{kodama1997}). ASTs in which small fragments are exchanged may go unnoticed. On the contrary, if two chromosomes break at their terminal part and form a dicentric chromosome, it will be clearly visible.

Another possibility is that the probability to form an ADS or an AST changes as the dose increase. Exchange type aberrations have a fast kinetics formation (\cite{darroudi1998}), and using fussion PCC techniques to evaluate the formation of dicentrics it was described that after 2 Gy irradiation the dicentrics were mostly formed during the first two hours post irradiation, while the resting unsolved damage seemed to reconstitute the original chromosomes (\cite{pujol2020}), it can be hypothetized that at higher doses the probability to form an ASD increases with respect the probability to form a AST. Consequently, further investigation and analysis are necessary to gain a better understanding of the underlying mechanisms and factors that influence the formation of dicentrics and translocations. The results observed in the present study have an implication in biological dosimetry studies. For past dose assessment, it is important to choose a specific biomarker, whether using the conventional nomenclature or PAINT modified one, and that the results obtained at high doses are not always interchangeable.

\section*{Ethical approval}

\section*{Disclosure statement}
No potential conflict of interest was reported by the authors.
\section*{Author contributions}
Conceptualization, D.M., P.P., JF.B.;
Methodology, D.M., P.P., C.A. and V.G.R.;
Software, D.M.;
Formal Analysis, D.M.;
Resources, JF.B. ;
Data Curation, JF.B.;
Writing--Original Draft, D.M. and JF.B.; Writing--Review\&Editing, D.M., P.P., C.A., V.G.R, JF.B.

\section*{Data and code availability}
All data and computer code used for analysis during this study are included in supplementary material.

\section*{Funding}
This work was supported by the Consejería de Educación, Cultura y Deportes (Junta de Comunidades de Castilla-La Mancha, Spain) [the Project MECESBayes, SBPLY/17/180501/000491]; Ministerio de Ciencia e Innovación (Spain) [research grants PID2019-106341GB-I00, PID2022-137414OB-I00]; the Spanish Consejo de Seguridad Nuclear [BOE-A-2019-311]; and the Spanish State Research Agency [the Severo Ochoa and Marıa de Maeztu Program for Centers and units of Excellence in R\&D (CEX2020-001084-M)].

\section*{Notes on contributors}
Dorota Mlynarczyk, PhD student, Mathematician, PhD student at the Department of Mathematics, Faculty of Sciences, Universitat Autònoma de Barcelona, Bellaterra (Catalonia), Spain. 

Pedro Puig, PhD, Mathematician, Professor at the Department of Mathematics and member of the Center de Recerca Matemàtica, Faculty of Sciences, Universitat Autònoma de Barcelona, Bellaterra (Catalonia), Spain.

Joan-Francesc Barquinero, PhD, Biologist, University Professor at the Department of Animal Biology, Plant Biology and Ecology, Faculty of Biosciencies, Universitat Autònoma de Barcelona, Bellaterra (Catalonia), Spain.

Carmen Armero, PhD, Mathematician, Professor at the Department of Statistics and Operations Research, Faculty of Mathematics, Universitat de València, Burjassot (Valencian Community), Spain.

Virgilio Gómez-Rubio, PhD, Mathematician, Professor at the Department of Mathematics, School of  Industrial Engineering, Universidad de Castilla-La Mancha, Albacete (Castilla-La Mancha), Spain.

\end{document}